\documentclass[11pt]{article}
\usepackage{epsfig}
\usepackage{amssymb}
\usepackage{a4} 
\newcommand{\be}{\begin{equation}}
\newcommand{\ee}{\end{equation}}
\newcommand{\ba}{\begin{array}}
\newcommand{\ea}{\end{array}}

\newcommand{\bea}{\begin{eqnarray}}  
\newcommand{\eea}{\end{eqnarray}}  

\newtheorem{prop}{Proposition}

\def\b#1{{\mathbb #1}}

\def\nn{\nonumber  \\}

\begin{document} 

 ~

~\\

\centerline{\bf \Large Towards soliton solutions of a perturbed sine-Gordon}
\centerline{\bf \Large equation} 
  
\bigskip
\medskip
\centerline{
A. D'Anna$^1$, M. De Angelis$^1$, G.
Fiore\footnote{Dip. di Matematica e Applicazioni, Fac.  di
Ingegneria,  Universit\`a di Napoli ``Federico II'', V. Claudio 21, 80125
Napoli. Email: danna@unina.it, modeange@unina.it,
gaetano.fiore@unina.it}~\footnote{I.N.F.N., Sezione di Napoli,
        Complesso MSA, V. Cintia, 80126 Napoli}} 

\date{}

\bigskip
~
\bigskip
\bigskip

{\small
\noindent
{\bf Abstract:} We give arguments for the existence of
{\it exact} travelling-wave solutions, of old (in particular solitonic) 
and new type, of a
perturbed sine-Gordon equation on the real line or on the circle, and classify
them. The perturbation of the equation consists of  a constant forcing term
and a linear dissipative term. Such solutions are allowed exactly by the
energy balance of these terms,  
 and can be observed experimentally e.g. in
the Josephson 
 effect in the theory of superconductors, which is one of 
the physical phenomena described by the equation. 
    
\medskip
}

\bigskip
\noindent  
Preprint 05-30 Dip. Matematica e Applicazioni, Univ. ``Federico II'',
Napoli\\
\noindent  
DSF/13-2005 
 \bigskip

\section{Introduction}

The ``perturbed'' sine-Gordon equation
  \be                               \label{equation}
\varphi_{tt}-\varphi_{xx}
    + \sin\varphi+\alpha \varphi_t+\gamma =0,
    \qquad x\in\b{R},   
 \ee
($\alpha\ge 0,\gamma\in \b{R}$ are constants)
 has been used to
describe with a good approximation a number of interesting physical phenomena,
notably Josephson effect in the theory of superconductors 
\cite{Jos}, or 
more recently also the propagation of localized
magnetohydrodynamic modes in plasma physics \cite{Sco04}. The last two
terms are respectively a dissipative and a forcing one; the (unperturbed) 
sine-Gordon equation is obtained by setting them equal to zero.
 
In the Josephson effect  
$\varphi(x,t)$ is the phase difference of the macroscopic quantum
wave-functions describing the Bose-Einstein condensates of
Cooper pairs in two superconductors separated by a very thin, narrow
and  long dielectric (a socalled ``Josephson junction'').
The $\gamma$ term is the (external) socalled ``bias current'',
providing energy to the system, whereas the dissipative term $\alpha
\varphi_t$ is due to Joule effect of the residual normal current across the
junction, due to electrons not paired in Cooper pairs.
Additional terms can be added
 to describe additional features, e.g. a term
like $\sigma \varphi_x$
 would approximately describe \cite{Napetal} the
Josephson effect in a junction with
 a linearly varying (albeit very small) 
breadth, whereas a dissipative term like $-\varepsilon \varphi_{xxt}$ would
take into account (see e.g. \cite{BarPat82}) the effect of the residual normal
current along the junction. We plan to address the latter, third order equation
elsewhere, exploiting our results \cite{DeaMonRen02,DanFio05} about equations
involving the differential operator 
$-\varepsilon\partial_x^2\partial_t+\partial_t^2-\partial_x^2$.

Among the solutions of the sine-Gordon equation the solitonic
ones are particularly important, in that they describe stable
waves propagating along the $x$-line. There are strong 
experimental (see e.g. \cite{BarPat82} for the Josephson
effect), numerical
\cite{Joh68,NakOnoNakSat74} and analytical
\cite{FogTrulBisKru76,McLSco77} indications that  
solutions of this kind are deformed, but survive in
the perturbed case; nevertheless, up to our knowledge there is no rigorous
proof of this. The analytical indications are obtained
within the by now standard perturbative method 
\cite{KauNew76,SatYaj74,KarMas77,KeeMcL77} based on
modulations of the unperturbed (multi)soliton solutions with slowly varying
parameters (tipically velocity, space/time phases, etc.…) and small radiation
components. (This is inspired by the Inverse Scattering Method).          
The Ansatz for the approximate one-(anti)soliton solution reads
$$
\varphi(x,t) =  \hat g_0\Big(x\!-\!x_0(t)\!-\!\tilde v(t)t\Big) \!+\!
\epsilon\varphi_1(x,t)\!+\!…... 
$$
where $\varphi_0(x,t):=  \hat g_0(x-vt)$ is one of the {\it unperturbed}
(anti)soliton solutions given below in (\ref{soliton}), whereas the slowly
varying $x_0(t),\tilde v(t)$  and the perturbative ``radiative'' corrections 
$ \epsilon\varphi_1(x,t)+…...$ have to be computed perturbatively in
terms    of the perturbation $\epsilon f$ of the sine-Gordon equation (in the
present case one may choose $\epsilon=\gamma$ and 
$\epsilon f=+\alpha \varphi_t+\gamma $). 
One finds in
particular approximate solutions with constant velocity
\be
                                          \label{v_atinf}
\tilde v(t)\equiv v_{\infty}:=\pm[1+(4\alpha/\pi\gamma)^2]^{-\frac 12}
\ee
which are characterized by a power balance between the dissipative term – 
$\alpha \varphi_t$
and the external force term –$\gamma$. They are interpreted as approximating
{\it expected} exact (anti)soliton solution.
 The experimentally
observed velocity is consistent with the value $v_{\infty}$ within
present experimental errors.

The purpose of this
 work is to give non-perturbative arguments for
the existence of {\it exact} travelling-wave (in particular solitonic)  
solutions of the
 above equation  on the real line or on the circle, and a
preliminary classification of them. Our
approach is less
 ambitious,  in that it is based on pushing forward the study
of the
 ordinary differential equation which is obtained  by replacing in the
equation (\ref{equation}) the standard travelling-wave Ansatz 
\be                            
\label{fixprof}  \varphi(x,t)=\tilde g(\tilde \xi)=\tilde g(x-vt)
\ee  
(here and in the sequel $\tilde \xi:=x-vt$),
and therefore cannot be applied to multisolitonic solutions. 
The ordinary
differential equation is the same as the one describing the motion along a 
line of a particle 
subject to a ``washboard'' potential and
immersed in a linearly viscous fluid, and therefore the
problem is essentially reduced to studying this simpler mechanical analog
where $\tilde \xi$ plays the role of `time'.
After reviewing (section \ref{SG}) travelling-wave solutions of the sine-Gordon
equation in section \ref{PSG} we classify the possible solutions of the
latter, identifying those having bounded  energy density at infinity and
possibly yielding stable solutions for (\ref{equation}); the main results are
collected   in Proposition \ref{Thm1}.  The
actual existence of such solutions is strongly suggested by our physical
expectations on the mentioned particle mechanical analog and will be proved
elsewhere \cite{Fio05s},  together with some general   
properties of these solutions (there we will also provide an alternative
perturbative scheme for their determination). 
Among the solutions there are:  those yielding solitonic
solutions for (\ref{equation}), which are characterized by their going to
two neighbouring local maxima of the potential energy of the particle as
$\tilde\xi\to\pm\infty$; those yielding ``array of (anti)solitons''
solutions; those yielding  ``half-array of (anti)solitons'' solutions.
The occurrence of the latter is a {\it new phenomenon}, with no counterpart in
the pure sine-Gordon case. Also, contrary
to the unperturbed case, the propagation velocity $v$ of the soliton turns out
to be not a free parameter, but a function of $\alpha,\gamma$, which coincides
\cite{Fio05s}, at lowest order in $\gamma$, with (\ref{v_atinf}).

\subsection{Preliminary considerations:}

Given a solution one can find a two-parameter family with infinitely many
others by
 space or time translations; therefore, for each family it suffices
to give just one representative element.

 \medskip

\begin{figure}[ht] 
\begin{center} 
\epsfig{file=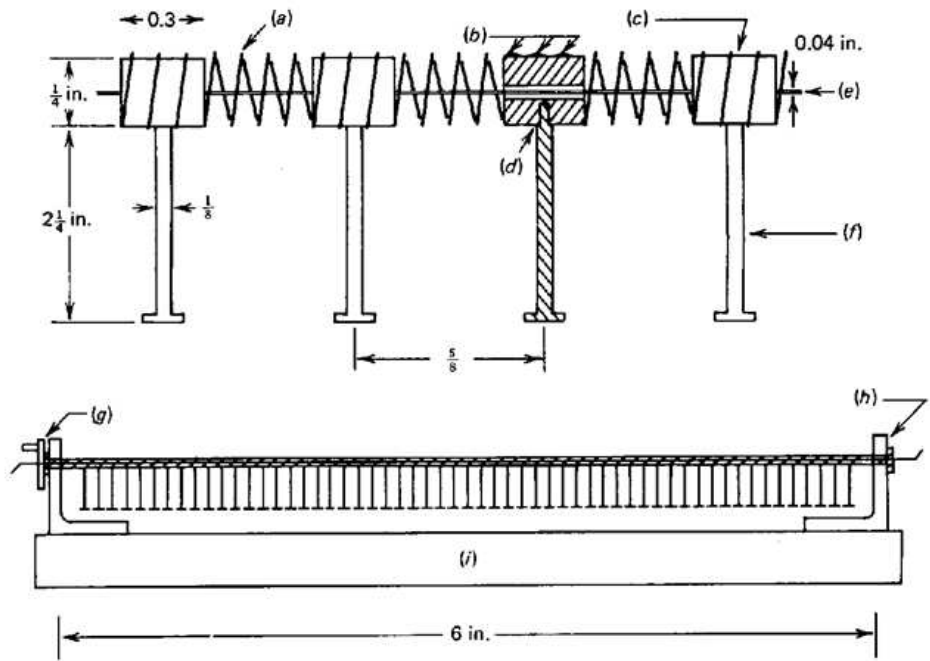 ,width=13truecm}  
\caption{Mechanical model for the sine-Gordon equation. (a) Spring, (b)
dolder, (c) brass, (d) tap and thread, (e) wire, (f) nail, (g) and (h) ball
bearings, (i) base (After A. C. Scott \cite{Sco70}, courtesy of A. Barone, see
\cite{BarPat82})} 
\label{PendulaChain}  
\end{center}  
\end{figure} 
A system obeying the sine-Gordon equation can be modelled \cite{Sco70} 
by the  discretized 
mechanical analog in fig. \ref{PendulaChain}, namely a chain of heavy
pendula constrained
 to rotate around a horizontal axis and coupled to each
other through a
 spring applying an elastic torque; one can model also the
dissipative term
 $-\alpha \varphi_t$ of
(\ref{equation}) by immersing the pendula in a linearly viscous fluid, and the forcing term
$\gamma$ by assuming that there is a uniform friction between the pendula and
the horizontal axis, and that the latter rotates with constant velocity.

\medskip
The constant solutions of (\ref{equation})
are $\varphi(x,t)\equiv - \sin^{-1}\gamma\!+\!2\pi k$
and $\varphi(x,t)\equiv  \sin^{-1}\gamma\!+\!(2k\!+\!1)\pi $. The former
are stable, the latter unstable. To see this one just needs to note 
that they yield respectively local minima and maxima of 
the energy density 
\be  
h:=\frac {\varphi_t^2}2+ 
\frac {\varphi_x^2}2+\gamma\varphi-\cos\varphi+K.
\ee 
This is visualized in the mechanical analog in fig. \ref{PendulaChain}
respectively by configurations with all pendula hanging down or standing up. 
We choose the free constant $K\in\b{R}$ so that it gives
a zero energy density at one (particular) stable constant solution,
$\varphi(x,t)\equiv - \sin^{-1}\gamma$: then
 $K=\sqrt{1-\gamma^2}+\gamma\sin^{-1}\gamma$. In general,
as a consequence of (\ref{equation}) $h$ fulfills the equation
\be                             \label{pertconteq} 
\partial_t h-\partial_x j=-\alpha\varphi_t^2.
\ee
where we have introduced the energy current density 
$j:=\varphi_x\varphi_t$. If  $\alpha=0$ this is a continuity equation.
The negative sign at the rhs
shows the dissipative character of the time derivative  term
in (\ref{equation}) if $\alpha>0$. 

Our {\it working definition of a (multi)solitonic solution} $\varphi$ 
is: A) It is a {\it stable solution which significantly
differs from some minima of the energy density $h$ only in some localized
regions}; this means that mod. $2\pi$ it must be   
\be           
         \label{asymcond}   \lim\limits_{x\to -\infty}\varphi(x,t)=-\sin^{-1}\!
\gamma,       \qquad  
 \lim\limits_{x\to +\infty}\varphi(x,t)=-\sin^{-1}\!\gamma+ 2n\pi 
\qquad
\ee 
with $n\in\b{Z}$. Moreover, as usual we require that: B) In the far past
$\varphi$  is approximately a superposition of single solitons, antisolitons
(and possibly breathers), which in the far future emerge from collisions again
with the same shape and velocities with which they entered [however, in this
work we shall only deal with single (anti)solitons]. As we shall recall below,
one-soliton and one-antisoliton solutions are characterized by $n=1,-1$
respectively [whereas the static stable and constant solution
$\varphi(x,t)\equiv \sin^{-1}\gamma \mbox{ (mod }2\pi)$ correspond to $n=0$]. 
In the mentioned mechanical analog the one-(anti)soliton solution describes a
localized twisting  of the pendula chain by $2\pi$ (anti)clockwise, as
depicted in figure \ref{Photographs} (a), moving with constant velocity. The
above condition yields an energy density $h$  (rapidly) going to $0,
2n\pi\gamma$ respectively as $x\to -\infty,\infty$; only if $\gamma=0$, i.e.
if the values of the  potential energy at all the minima coincide, the
potential energy $h$ vanishes at both $x\to -\infty,\infty$, and one 
recovers the standard definition of solitons.

Although $n\neq 0$ makes the total Hamiltonian 
\be  
H:=\int\limits_{-\infty}^{+\infty}h(x,t)dx 
\ee 
divergent, it gives a well-defined, non-positive time-derivative
$$
\dot
H=-\int\limits^{\infty}_{-\infty}\alpha\varphi_t^2
\le 0,
$$
as a result of integration of (\ref{pertconteq}). The effect of $\gamma\neq 0$
is to make the values of the energy potential at any two minima different;
this leaves room for an indefinite compensation of the dissipative power
loss by a falling down in the total potential energy, and so may account for
solutions not being damped to constants as $t\to\infty$.

\medskip
Without loss of generality we can assume $\gamma \ge 0$. If originally
this is not the case, one just needs to replace $\varphi\to -\varphi$.
If $\gamma> 1$ no solutions $\varphi$ having 
finite limits and vanishing derivatives for $x\to \pm \infty$  can exist, in particular
no static solutions. If $\gamma=1$ the only static solution $\varphi$ having for
$x\to \pm \infty$  finite limits and vanishing derivatives is 
$\varphi\equiv -\pi/2\mbox{ (mod }2\pi)$, which however is unstable. 
In the sequel we shall assume $0\le \gamma<1$.

\section{The sine-Gordon equation}
\label{SG}

Before going on, let us recall \cite{BarEspMagSco71,ChuMcLSco73} what happens  
in the case $\gamma=0=\alpha$. 
The sine-Gordon equation is
\be                          
\varphi_{tt}-\varphi_{xx} + \sin\varphi =0,          \label{sinegor}
  \qquad   \qquad x\in\b{R}.   
\ee
The associated Hamiltonan (a conserved quantity) is
\be 
H_0:=\int\limits_{-\infty}^{+\infty}\left(\frac {\varphi_t^2}2+
\frac {\varphi_x^2}2-\cos\varphi+1\right)dx.
\ee

Let $\tilde \xi:=x-vt$. One looks for travelling-wave solutions
of (\ref{sinegor}), namely for solutions with a `fixed-profile' Ansatz 
(\ref{fixprof}). As the time $t$
flows the profile of $\tilde g$ will move to the right/left according to a
positive/negative sign of $v$. Replacing the Ansatz  in the sine-Gordon 
equation one finds for $\tilde g$ the equation 
$(v^2-1)\tilde g''+\sin \tilde g=0$.
If $v^2=1$ the equation admits only the 
(stable or unstable) static solutions $\varphi(x,t)=\tilde
g(\tilde\xi)\equiv  0,\pi$ (mod $2\pi$). Otherwise
it can be rewritten in the form  
\be
g''+\sin  g=0,
\ee 
where
\be
\xi=\sigma\frac{\tilde\xi}{\sqrt{|v^2\!-\!1|}},\qquad 
g(\xi):=\eta\cases{\tilde g(\xi) \:\:\:\mbox{ if
}v^2\ge 1 \cr \tilde
 g(\xi)\!-\!\pi \:\:\:\mbox{ if }v^2< 1,}                
   \label{redef}
\ee
where $\eta,\sigma=\pm 1$ are arbitrary signs. This is the equation of
motion w.r.t. the `time' $\xi$ of a pendulum ($g$ being the deviation
angle from the stable equilibrium position), or
equivalently of a particle in a sinusoidal ``energy potential'' 
$U=-\cos g$.
Multiplying the equation by $g'$ one finds the ``mechanical energy 
of the pendulum'' integral of motion {\rm e}
\be
\qquad \frac{d}{d\xi} \left[\frac{g'{}^2}2 \!-\!\cos g\right]=0
\qquad \Rightarrow \qquad  
\mbox{\rm e}:=\frac{g'{}^2}2\!+\!U=\mbox{const}. 
\ee
By definition $\mbox{\rm e}\ge -1$. The above implies
$$
\pm dg\frac{1}{\sqrt{2(\mbox{\rm e}\!+\!\cos g)}}=d\xi.
$$
The latter equation can be immediately integrated out, giving
$\xi=\xi(g)$. The final step is the inversion of this
function in disjoint intervals, which will give $g=g(\xi)$, and patching
the intervals.

\begin{figure}
\begin{center}
\epsfig{file=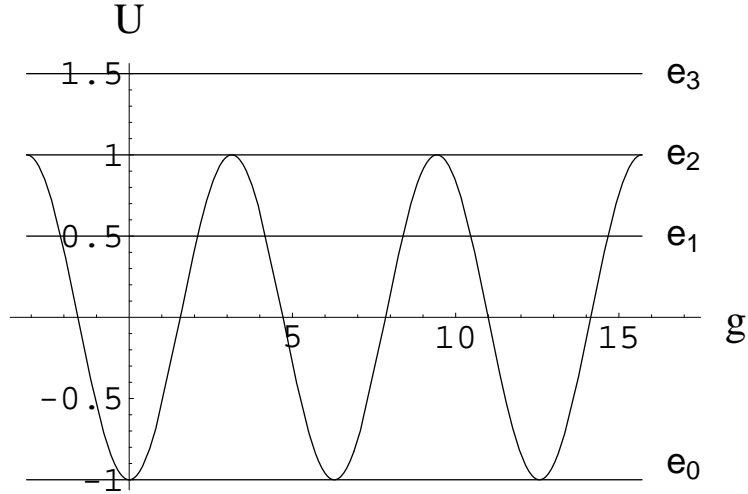,width=11truecm} 
\caption{The potential energy $U$}
\label{figura1}
\end{center}
\end{figure}
According to the choice of $\mbox{\rm e},v$ one obtains different kinds 
of solutions.
Plotting the potential energy (Fig. \ref{figura1}) helps us get an immediate 
qualitative understanding of them.

\begin{figure}
\begin{center} 
\epsfig{file=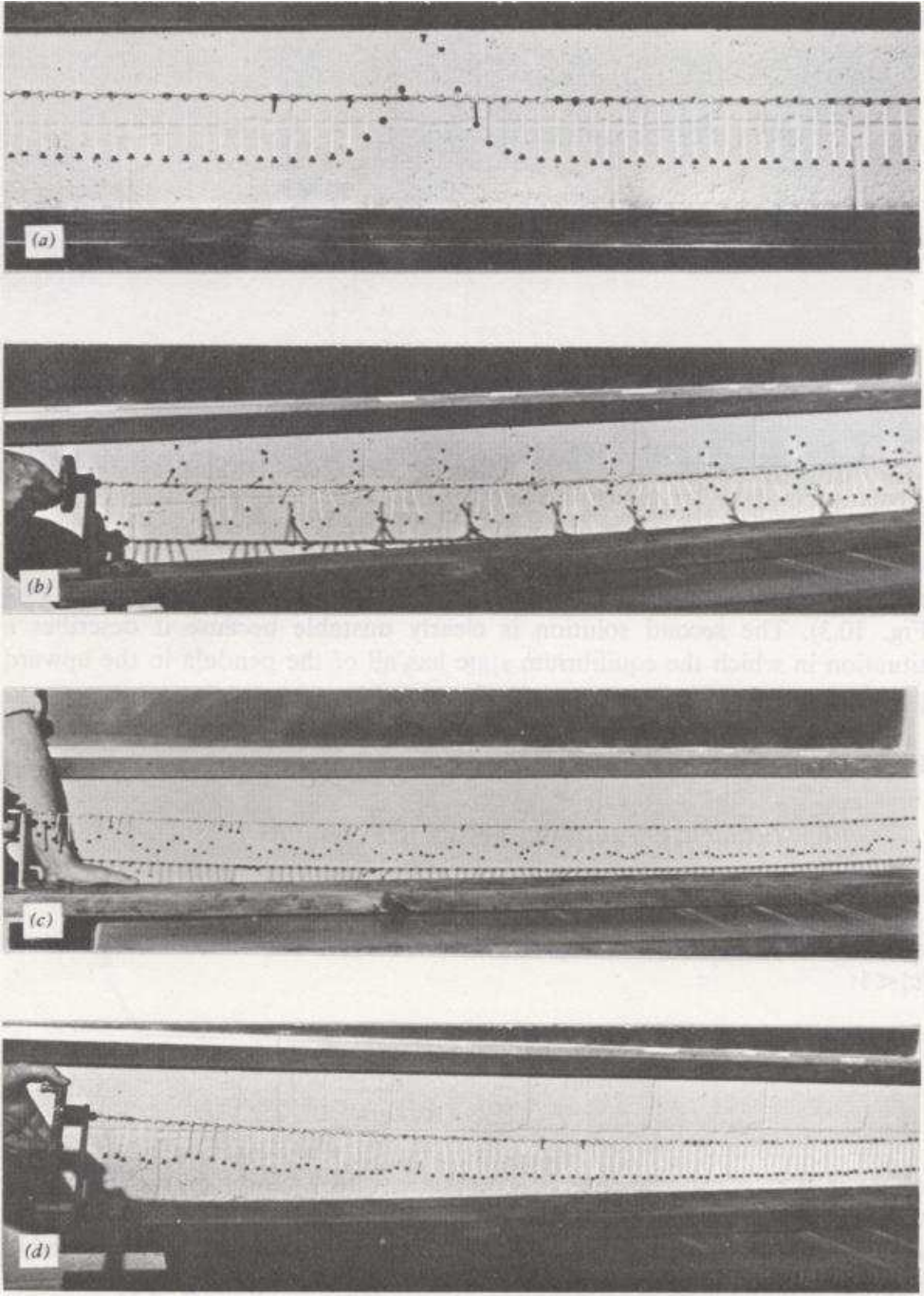,width=13truecm}  
\caption{Photographs of the mechanical model of fig. \ref{PendulaChain}: (a)
single soliton solution, (b) evenly spaced array of solitons, large (c) and
small (d) amplitude plasma wave solutions (After A. C. Scott
\cite{Sco70}, courtesy of A. Barone, see  \cite{BarPat82})}  
\label{Photographs}   
\end{center}  
\end{figure}

\begin{enumerate}

\item If $\mbox{\rm e}=\mbox{\rm e}_0:=-1$ then necessarily $g(\xi)\equiv
0\,(\mbox{mod}2\pi)$. This corresponds to the  constant solutions
\be
\varphi^s(x,t)\equiv 0\,(\mbox{mod}2\pi)\qquad\qquad
\varphi^u(x,t)\equiv\pi\,(\mbox{mod}2\pi)                                     
             \label{constunpert} \ee
respectively if  $v^2> 1$ and  $v^2<1$. The former (resp. latter) is
clearly stable (resp. unstable) because it corresponds to all pendula 
hanging downwards (resp. standing upwards) in the  model of fig.
\ref{PendulaChain}. 
 The same constant solutions
arise from considering the constant $g=(2k+1)\pi$, corresponding to
$\mbox{\rm e}=\mbox{\rm e}_2=1$.

\item If $-1<\mbox{\rm e}<1$ ($\mbox{\rm e}=\mbox{\rm e}_1$ in Fig.
\ref{figura1}) then the corresponding solution $\bar g_0(\xi;\mbox{\rm
e})$ can be written in terms of elliptic
functions and describes the motion of a particle confined in an interval
contained in $]-\pi,\pi[$ and oscillating around $g=0$
($\mbox{mod}2\pi$) with some period $\Xi_p(\mbox{\rm e})$.
For $\varphi$, this translates into
a periodic oscillating wave travelling with velocity $v$:
If $v^2> 1$ then $\varphi$ oscillates around the stable equilibrium
solution $\varphi^s\equiv 0$ and describes a ``plasma wave'', see fig.
\ref{Photographs} (c), (d);  if $v^2< 1$ then
$\varphi$ oscillates around the unstable equilibrium  solution
$\varphi^u\equiv \pi$.  Both kinds of $\varphi$ are however
unstable \cite{Sco69',BarEspMagSco71,ChuMcLSco73}.

\item If $\mbox{\rm e}=\mbox{\rm e}_2=1$, beside the constant
solution yielding (\ref{constunpert}), there are in addition the
solutions
$$
\hat g_0(\pm \xi)= 4 \tan^{-1}\left[\exp(\pm \xi)\right] -\pi.
$$
 Mod $2\pi$, $\hat g_0(\xi)\to \pm\pi$ as $\xi\to
\pm\infty$: the particle, confined in the interval
$-\pi\!<\!g\!<\!\pi$, starts at `time' $\xi=-\infty$ from one top of
the energy potential and reaches the other one at $\xi=\infty$.  Mod.
$2\pi$, they translate into the following solutions of the original
problem (\ref{sinegor}) 
\bea 
&&\varphi_0^{\pm}(x,t;v) = 4\tan^{-1}\left\{\exp\left[\pm
\frac{x-vt}{\sqrt{v^2\!-\!1}}\right]\right\}-\pi \qquad \mbox{ if
}v^2> 1,\qquad \nonumber\\
&&\hat \varphi_0^{\pm}(x,t;v) =  4 \tan^{-1}\left\{\exp\left[\pm
\frac{x-vt}{\sqrt{1\!-\!v^2}}\right]\right\}\qquad \mbox{ if }v^2<
1 .\qquad              \label{soliton}
\eea
Both families represent solutions  (rotating
clockwise or anti-clockwise according to the sign $\pm$) with localized region
of variation of
$\varphi$ and travelling
with velocity $v$. The first families are clearly unstable because 
they
describe solutions of the  model of fig. \ref{PendulaChain} where all 
pendula stand upwards except in the small moving region where
they twist once around the axis;
 their values of the total mechanical energy
$H$ are infinite.
 The second families are stable \cite{Sco69',BarEspMagSco71,ChuMcLSco73}, 
as it can be expected from the fact that they describe solutions of the
mechanical model of fig. \ref{PendulaChain} where all 
pendula hang downwards except in the small moving region where
they twist once around the axis, see fig.
\ref{Photographs} (a);
their values of the total mechanical energy $H$ are finite.
 They represent
respectively a soliton ($\varphi_0^+$) and an antisoliton
($\varphi_0^-$) travelling with velocity $v$. Note that they
fulfill
 (\ref{asymcond}) with $n=\pm 1$. For $v=0$ we have a static
(anti)soliton.

\item If $\mbox{\rm
e}=\mbox{\rm e}_3>1$ (see Fig.  \ref{figura1}) then the
corresponding solution $\check g_0(\pm \xi;\mbox{\rm e})$ describes
a particle moving towards the right and the left respectively, `for
ever', since it has a sufficient energy to overcome the tops of the
energy potential.  Moreover, its kinetic energy and velocity are
periodic with some period $\Xi_0(\mbox{\rm e})$, (with
$\Xi_0\to\infty$ as $\mbox{\rm e}\downarrow 1$). This means, for any
$\xi\in\b{R}$,  
\be \check g_0(\xi+\Xi_0;\mbox{\rm e})=\check g_0(\xi;\mbox{\rm e})+2\pi,
\label{mod2piperiodic} \ee i.e. $g(\xi)$ is the sum of a linear and of a periodic
function.  Again, the
corresponding solutions of the original problem (\ref{sinegor}) are
unstable or \cite{Sco69',BarEspMagSco71,ChuMcLSco73} stable
according to $v^2\!> \!1$ or $v^2\!<\! 1$,   because they correspond
to `most' pendula up or down in the model of fig.
\ref{PendulaChain}.  The stable solutions  ($v^2\!< \!1$)
$\check\varphi_0^{\pm}$  describe  evenly spaced ``arrays of
solitons and antisolitons'', travelling with velocity $v$, see fig.
\ref{Photographs} (b).

\end{enumerate}

No $n$ with $|n|>1$ is allowed for travelling wave solutions fulfilling
(\ref{asymcond}).

\section{Adding the forcing and the first order dissipative terms  in the
equation}

\label{PSG}
We still adopt the Ansatz (\ref{fixprof}) for the solutions
of (\ref{equation})   and
ask how the classes of solutions found in the preceding 
section are deformed. We are interested
in stable solutions $\varphi$ with bounded derivatives for 
$x\to \pm \infty$.
Having  in mind the mechanical analog of the chain of pendula, it is natural
to expect that solitonic solutions will survive perturbation: if we just 
smoothly switch $\gamma$  then even the static soliton   
will start to move and gradually  accelerate; if we now also switch
$\alpha$,  we expect that its propagation will approach a steady regime 
in which dissipation and forcing balance  each other. 
We are actually going to see 
that not only the solitonic, but also the other classes of solutions found in
the previous section survive perturbation whenever a compensation of the 
forcing term $\gamma$ with the dissipation term
 in $\alpha$ can take place.

Replacing the Ansatz (\ref{fixprof}) in (\ref{equation}) 
we find the equation
\be                               \label{equation'}
    (v^2-1)\tilde g''-\alpha v \tilde g'
    + \sin \tilde g+\gamma=0,
    \qquad x\in\b{R}.
\ee
If $v=\pm 1$ the second order derivative in (\ref{equation'})
disappears and we get
$$
\pm\,\alpha \tilde g' = (\sin \tilde g+\gamma)  \label{equa} .
$$
Unless $\tilde g$ is constant (and therefore equal to 
$-\sin^{-1}\gamma$ or $\pi\!+\!\sin^{-1}\gamma$) then
there exists a $\tilde\xi_0$ such that $\tilde g'(\tilde\xi_0)\neq 0$ and 
$\tilde g_0:=\tilde g(\tilde\xi_0)\neq -\sin^{-1}\gamma,\pi+\sin^{-1}\gamma$. 
Integrating in a neighbourhood of $\tilde\xi_0$ one finds
$$
\tilde\xi-\tilde\xi_0=\int\limits^{\tilde\xi}_{\tilde\xi_0}d\tilde\xi'=
\pm \alpha\int\limits^{\tilde
g}_{\tilde g_0} \frac{dz}{\sin z+\gamma}.
$$ 
As $\tilde g$ approaches respectively $-\sin^{-1}\gamma$ or
$\pi\!+\!\sin^{-1}\gamma$ (mod. $2\pi$) the denominator goes to zero linearly
while keeping the same sign, and therefore the integral  diverges
logarithmically, implying that the corresponding time $\tilde\xi$ goes
respectively to $\pm \infty$ (or viceversa).  The corresponding solution
$\varphi$ for (\ref{equation})
 is unstable, therefore is not interesting for
our scopes, because:  it yields a maximum of the energy density $h$ 
as either $\tilde\xi\to\infty$, or
$\tilde\xi\to -\infty$, i.e. there it corresponds
to all pendula standing upwards  (while hanging downwards resp. as $\tilde\xi\to
-\infty$ or $\tilde\xi\to\infty$).

\medskip
If $v^2\neq 1$
we perform the redefinitions [compare with (\ref{redef})] 
\be
\ba{ll}
\xi:=-\mbox{sign}(v)\frac{\tilde \xi}{\sqrt{v^2\!-\!1}}\qquad 
& g(\xi)\!:=\!-\tilde g(\xi) \:\:\:\mbox{ if }v^2\!> \!1, \\[8pt] 
\xi\!:=\!\mbox{sign}(v)
 \frac{\tilde \xi}{\sqrt{1\!-\!v^2}}
\qquad & g(\xi)\!:=\!\tilde g(\xi)\!-\!\pi
\:\:\:\mbox{ if }v^2\!<\! 1,\\[8pt] 
 
\begin{figure}
\begin{center}
\epsfig{file=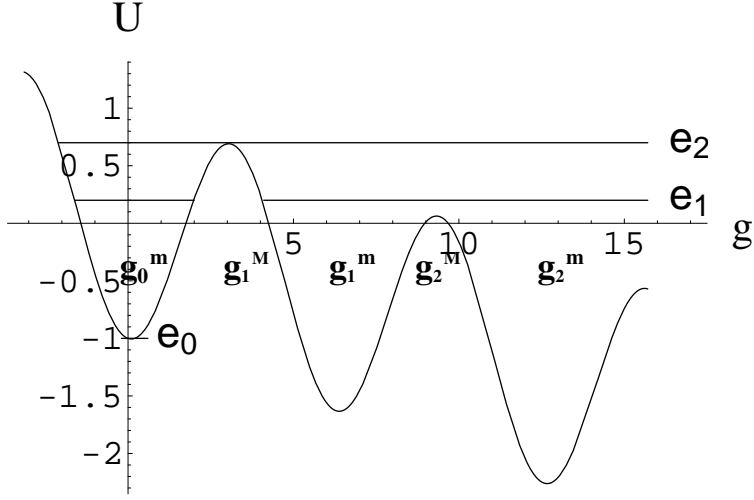,width=11truecm} 
\caption{The potential energy $U$ for  $\gamma=.1$
and three mechanical-energy levels for $\mu=0$ (no dissipation).}
\label{figura2}
\end{center}
\end{figure}

\xi\!:=
\!
\tilde \xi\!=\!x\qquad & g(\xi)\!:=\!\tilde g(\xi)-\pi \:\:\:\mbox{ if }v=0;
\ea              \qquad                    \label{redef'} 
\ee
we obtain
\be                               \label{equation"}
g''+\mu g'+U_g(g)=0,
    \qquad \xi\in\b{R},  
\ee 
which can be regarded as the 1-dimensional equation of
motion  w.r.t. the `time' $\xi$ of a particle with unit mass, 
a `potential energy'
$U(g)=-(\cos g+\gamma g)$ (see Fig. \ref{figura2}) and a
viscous force with viscosity coefficient 
\be                                       \label{defmu} 
\mu:=\frac{\alpha}{\sqrt{|v^{-2}-1|}} 
\ee 
[in other words, in the equation $\alpha, v$ appear 
only through their combination (\ref{defmu})]. 
One immediately finds
that the `mechanical energy' is not `time' independent,
but decreases according to
\be
\mbox{\rm e}'= -\mu g'{}^2\le 0               \label{edecrease}
\ee
$U$ admits local minima (resp. maxima) in the points 
$$
g_k^m:=\sin^{-1}\gamma\!+ \!2k\pi\qquad \qquad(\mbox{resp. }
g_k^M:=\pi\!-\!\sin^{-1}\gamma\!+\!2k\pi),
$$ 
and their corresponding values of
$U$  are 
\bea
&&U(g_k^m)=-\left[\gamma(\sin^{-1}\gamma+2k\pi)+\sqrt{1-\gamma^2}\right],\nn
&&U(g_k^M)=-\left[\gamma\left(-\sin^{-1}\gamma+(2k\!+\!1)\pi\right)-
\sqrt{1-\gamma^2}\right];\nonumber
\eea
they linearly decrease with $k$.

We have to look for solutions
of (\ref{equation"}) such that $g'$ is bounded all over
$\b{R}$. 
Multiplying (\ref{equation"}) by $g'$ and integrating over an interval
$[\xi,\bar\xi]$ we find
\be                               \label{integrel}
\qquad \frac 12 g'{}^2(\xi)=\frac 12 g'{}^2(\bar\xi)+\left[U\Big(g(\bar\xi)\Big)
\!-\!U\Big(g(\xi)\Big)\right]
+\mu\int\limits^{\bar\xi}_{\xi} g'{}^2(\xi')d\xi'.
\ee

\begin{enumerate}

\item The constant (in $\xi$) solutions corresponding to the local 
minima, maxima of the potential energy become $g(\xi)\equiv g_k^m$, $g(\xi)\equiv 
g_k^M$. The latter are the solutions for which there exists a $\bar\xi$ such 
that  $g(\bar\xi)=g_k^M$ and $g'(\bar\xi)=0$ (actually this happens for all 
$\bar\xi\in\b{R}$).  The particular ones $g(\xi)\equiv g_0^m$, $g(\xi)\equiv 
g_0^M$ correspond to the energy levels ${\rm e}= {\rm e}_0:=U(g_0^m)$, ${\rm 
e}= {\rm e}_2:=U(g_0^M)$ in Fig. \ref{figura2}. 
These solutions translate into the following 
constant solutions for the original problem:  (mod. $2\pi$)
\be 
\qquad\varphi^s(x,t)\!\equiv\! -\!\sin^{-1}\!\gamma
\qquad \varphi^u(x,t)\!\equiv\! \sin^{-1}\!\gamma\!+\!\pi
\ee 
(the former are stable, the latter unstable). 
 
\item If there exists some `time' $\bar\xi$ such that 
$g(\bar\xi)=g_k^M$ and $g'(\bar\xi)<0$
(particle located at some maximum of $U$ and moving leftwards), then
necessarily\footnote{In fact, since $U(g_k^M)>
U(g)$ for all $g>g_k^M$, each term at the rhs of (\ref{integrel}) is
positive for all $\xi<\bar\xi$, implying that $|g'(\xi)|>|g'(\bar\xi)|>0$ for
all $\xi<\bar\xi$. Hence the first limit in the previous relation. Now for
$g(\xi)$ growing without bound as $\xi\to -\infty$ the local maxima of 
$U\Big(g(\xi)\Big)$ decrease linearly, implying at least a linear growth
of the rhs of (\ref{integrel}), whence the second limit.}
$$
g(\xi)\stackrel{\xi\to -\infty}{\longrightarrow}\infty,\qquad\qquad\qquad
g'(\xi)\stackrel{\xi\to -\infty}{\longrightarrow}-\infty,
$$
so for our scopes we can readily exclude this case. 

\item If there exists no  `time' $\bar\xi$ such that 
$g(\bar\xi)$ is a local maximum of $U$, then necessarily the  motion is
confined in some interval $]g^M_{k\!-\!1},g^M_k[$ and $|g'(\xi)|$ by
(\ref{integrel}) keeps
 bounded. We have
already treated the constant $g(\xi)\equiv g_k^m$ solutions,
 so let us
consider the others.
 
 If $\mu=0$ (note that this is the case not only if $\alpha=0$, but
also if $\alpha>0$ and $v=0$, i.e. for a static $\varphi$) and:  
\begin{enumerate}
\item  ${\rm e}< U(g_k^M)$, then $g(\xi)$
will be periodic, oscillating `for ever' around $g_k^m$; in fig. \ref{figura2}
this corresponds to {\rm e}={\rm e}$_1$ 
 with $g\in]g^M_{-1},g^M_0[$.
The corresponding 
solutions $\varphi$ of the original problem will describe again 
unstable oscillations around the unstable equilibrium 
position if $v^2<1$, and  unstable
`plasma wave' oscillations around the stable equilibrium 
position if $v^2>1$.

\item   ${\rm e}=U(g_k^M)$ , then  
$g(\xi)\to g_k^M$ for both $\xi\to\pm\infty$; in fig.
\ref{figura2} this corresponds to  {\rm e}={\rm e}$_2$  with
$g\in]g^M_{-1},g^M_0[$. If $v^2>1$ this will yield
again an unstable solution $\varphi$, because the latter corresponds to 
all the pendula standing upwards at infinity, whereas if $v^2<1$ this
will yield a new type of solution $\varphi$ of the original problem, a 
kind of rigidly bounded soliton-antisoliton pair.
Whether this is stable or not should be investigated.

 \end{enumerate}

If $\mu$ (and therefore $\alpha$) is positive, then 
necessarily\footnote{The second limit follows from
(\ref{edecrease}) and the existence of the lowest bound $U(g_k^m)$ for
$\mbox{\rm e}$, using standard methods in stability theory. As for the first
limit, note that $I(\xi):=\int\limits^{\bar\xi}_{\xi} g'{}^2(\xi')d\xi'$
is nonnegative and monotonic. If {\it per absurdum}   $I(\xi)$ did not converge
to a finite $I_M>0$ but diverged as $\xi\to
-\infty$, then so would do the rhs(\ref{integrel}) (because the term in square
bracket is bounded for $g\in]g^M_{k\!-\!1},g^M_k[$), the lhs and hence $|g'|$,
in contradiction
with  the motion being confined in the interval. 
$I(\xi)\stackrel{\xi\to -\infty}{\longrightarrow} I_M$ implies 
{\it a fortiori} the first limit, as claimed.}
$$
g'(\xi)\stackrel{\xi\to -\infty}{\longrightarrow}0,
\qquad\qquad g'(\xi)\stackrel{\xi\to \infty}{\longrightarrow}0.
$$
As a
consequence, using the equation of motion (\ref{equation"}), as
$\xi\to\pm\infty$ $g$ must go to one of the following values:
$g^M_{k\!-\!1},g_k^m,g_k^M$. If in addition there exists some  `time'
$\bar\xi$ such that $g'(\bar\xi)<0$ (particle moving leftwards), then
necessarily\footnote{The first limit follows from excluding $g_k^m$, which would imply 
the constant $g(\xi)\equiv g_k^m$ solution. As for the second limit,  we can
readily exclude $g^M_{k\!-\!1}$ by the decrease law  (\ref{edecrease}); 
we have to exclude $g_k^M$ by the same law, if 
$g(\xi)\stackrel{\xi\to -\infty}{\longrightarrow}g_k^M$, and
by the existence of $\bar\xi$, if 
$g(\xi)\stackrel{\xi\to -\infty}{\longrightarrow}g^M_{k\!-\!1}$.}
$$ 
g(\xi)\stackrel{\xi\to -\infty}{\longrightarrow}g_k^M\mbox{ or
} g^M_{k\!-\!1},\qquad\qquad\qquad g(\xi)\stackrel{\xi\to
\infty}{\longrightarrow}g_k^m. 
$$
Such a $g$ translates into an unstable solution $\varphi$ of the original
problem, because it yields a maximum of the energy density $h$, i.e. it corresponds
to all pendula standing upwards, for either $\xi\to\infty$ 
or $\xi\to-\infty $ [depending which of the two possible definitions of
$g$ in (\ref{redef}) is adopted].

We are left with the last, most interesting possibility, namely a solution
$\hat g$ such that
\be                                  \label{solasym}
\hat g(\xi)\stackrel{\xi\to -\infty}{\longrightarrow}
g^M_{k\!-\!1}\qquad\qquad\qquad \hat g(\xi)\stackrel{\xi\to
\infty}{\longrightarrow}g_k^M
\ee
and $\hat g'(\xi)>0$ for all $\xi$.
This will translate into an unstable solution $\varphi$ (almost all pendula
standing upwards) if $v^2>1$ [see (\ref{redef})],
candidate stable solutions $\hat\varphi^{\pm}$ (almost all pendula
hanging downwards) if $v^2<1$. The latter  will describe the propagation of an
(anti)soliton with velocity $v$ [fig.
\ref{Photographs} (a)].
That such a solution exists can be argued as follows. 
Impose just  (\ref{solasym})$_1$. Consider first the case
$\mu=0$: the mechanical energy of the particle has the constant value
${\rm e}=U(g^M_{k-1})$, the corresponding solution will reach and overcome
$g_k^M$ after a finite time. Second, it is easily expected and not difficult to
show \cite{Fio05s} that for sufficiently large $\mu$ the corresponding solution will
never reach $g_k^M$, but rather invert its motion at some $g< g_k^M$ and go to
$g_k^m$ as $\xi\to\infty$. By continuity, we expect that there exists a
special value $\hat\mu(\gamma)$ such that the corresponding solution
$\hat g$ fulfills also (\ref{solasym})$_2$. As an immediate consequence
of the definition of $\mu$ the propagation velocity $v$ (for
$v^2<1$) is determined [see formula (\ref{solvel}) below]. One can determine
$\hat\mu(\gamma)$ perturbatively in $\gamma$ \cite{Fio05s}; at
lowest order one finds 
$\hat\mu(\gamma)=\pi\gamma/4+...$, which gives a $v$ coinciding
with (\ref{v_atinf}), but higher orders will correct the latter formula.

\item It remains to consider the cases  that there exists some `time' $\xi_k$
such that  $g(\xi_k)=g_k^M$
(particle located at some maximum of $U$), 
but $g'(\xi_k)>0$ (particle moving rightwards) at all such `times'. Then necessarily 
$g'(\xi)>0$ for all (finite) $\xi<\xi_k$.\footnote{Otherwise, {\it per
absurdum}  denote by
$\bar\xi$ the largest $\bar\xi<\xi_k$ such that $g'(\bar\xi)=0$. Then by 
(\ref{equation"}) it is
necessarily $-U_g\Big(g(\bar\xi)\Big)=g''(\bar\xi)\ge 0$. Since it cannot be
$U_g\Big(g(\bar\xi)\Big)=0$, it must necessarily be 
$g''(\bar\xi)>0$, what implies $g'(\xi)<0$ in a left neighbourhood
of $\bar\xi$ (namely $\bar\xi$ is a time of inversion of the motion); by
(\ref{edecrease}), going further backwards in time ${\rm e}$ will not decrease,
and therefore
there will be a $\xi_k'<\bar\xi<\xi_k$ such that 
 $g(\xi_k')=g_k^M$ and
$g'(\xi_k')<0$.} As a consequence, it will be
 
\be       \label{interest2}
 \ba{llll}
\mbox{either }\:\:
 &g(\xi)\stackrel{\xi\to
-\infty}{\longrightarrow}-\infty
\qquad &\mbox{and}\qquad
& g(\xi)\stackrel{\xi\to \infty}{\longrightarrow} \infty \\[8pt] 
\mbox{or }\:\: &g(\xi)\stackrel{\xi\to -\infty}{\longrightarrow} 
g^M_{l}\qquad &\mbox{and}\qquad &g(\xi)\stackrel{\xi\to
\infty}{\longrightarrow} \infty 
\ea
\ee
with some $l<k$, with $|g'(\xi)|$ remaining bounded.

Consider first the case
(\ref{interest2})$_1$. Employing as before a continuity argument in
$\mu$ and the invariance of Eq. (\ref{equation"}) under $g\to g+2\pi$,
we actually infer that for any $g'_0>0$ there exists
a special value $\check{\mu}(\gamma,g'_0)$ and a finite
time $\check\Xi(\gamma,g'_0)$ such that the
corresponding solution  $\check g(\xi;g'_0,\gamma)$ fulfills
$$
\ba{l}
\check g(0;g'_0,\gamma)=g_{k-1}^M\qquad \check g'(0;g'_0,\gamma)=g'_0\\
\check g(\check\Xi;g'_0,\gamma)=g_k^M\qquad \check
g'(\check\Xi;g'_0,\gamma)=g'_0 \ea
$$
and consequently, more generally,
\be
\check g(\xi+\Xi)=\check g(\xi)+2\pi.
\ee
Since, as one can expect and as we shall prove in \cite{Fio05s},
the dependence of $\Xi$ on $g'_0$ is strictly monotonic, one can
choose $\Xi$ instead of  $g'_0$ as an independent variable.
Again, this may translate into stable solutions $\check \varphi$
for (\ref{equation}) only if $v^2<1$, and, as an immediate
consequence of the definition of $\mu$, the propagation velocity $v$
is determined [see formula (\ref{arrvel}) below].
Therefore $\check \varphi$ will describe an array
of (anti)solitons travelling with such a velocity [fig.
\ref{Photographs} (d)].
The (anti)soliton solutions $\hat \varphi$ can also be regarded and obtained
as $g'_0\downarrow 0$ limits of $\check \varphi$. 

As for the cases (\ref{interest2})$_2$, we have hints 
\cite{Fio05s} that the corresponding
$g$ will approach a $\check g$ as $\xi\to\infty$.

\end{enumerate}

We collect our main results in
\begin{prop}
Mod.  $2\pi$, stable, travelling-wave solutions of (\ref{equation}) 
(where $1>\gamma>0$ and $\alpha\ge 0$) having
bounded derivatives at infinity can be only of the following types (with 
$\xi:=(x\!-\!vt)/\sqrt{|1\!-\!v^2|}$):
\begin{enumerate}
\item The static solution 
$\varphi^s(x,t)\equiv -\sin^{-1}\gamma$.

\item  The soliton/antisoliton solutions 
$\hat\varphi^{\pm}(x,t)=\hat g(\pm \xi)$
\be
\lim\limits_{x\to -\infty}\hat\varphi^{\pm}(x,t)=-\sin^{-1}\gamma
\qquad\quad
\lim\limits_{x\to \infty}\hat\varphi^{\pm}(x,t)=-\sin^{-1}\gamma\pm 2\pi
\ee
travelling rightwards/leftwards resp. with velocity
\be 
v=\pm \frac{\hat\mu(\gamma)}{\sqrt{\alpha^2+
\hat\mu^2(\gamma)}}\equiv \pm\hat v;                \label{solvel}
\ee
at lowest order in $\gamma$ 
the function $\hat\mu$ is given \cite{Fio05s} by 
$\hat\mu(\gamma)=\pi\gamma/4+...$.

\item For any period $\Xi\in]0,\infty[$
the ``arrays of solitons/antisolitons'' solutions 
$\check\varphi^{\pm}(x,t)=\check g(\pm \xi)$, where 
\be
\check g(\xi+\Xi)=\check g(\xi)+2\pi
\ee
travelling rightwards/leftwards resp. with velocity
\be 
 v=\pm \frac{\check{\mu}}{\sqrt{\alpha^2+
\check{\mu}^2}}\equiv \pm\check v.               \label{arrvel}
\ee
$\check{\mu}$ 
(as well as $\check g$ itself) is a function of  $\gamma, \Xi$.

\item  The ``half-array of solitons/antisolitons''
solutions  $\varphi^{\pm}(x,t)= g(\pm \xi)$
\be
\lim\limits_{\xi\to -\infty}g(\xi)=-\sin^{-1}\gamma
\qquad\quad 
\lim\limits_{\xi\to \infty}g(\xi)=\infty.
\ee
trevelling rightwards/leftwards resp. with velocity $\pm\check v$.

\end{enumerate}
\label{Thm1}
\end{prop}

{\bf Remark 1.}
We emphasize that, in contrast with the unperturbed soliton (and array of 
solitons) solutions, where $v$ was a free parameter (of modulus less than 1), 
$v$ for the corresponding perturbed soliton (and array of 
solitons) solutions is predicted as the function of 
$\gamma,\alpha, g'_0$ given by formulae (\ref{solvel}), (\ref{arrvel}).

{\bf Remark 2.}
Note that $\check g$ is a solution of (\ref{equation"}) also if we
identify $\xi+\Xi$ with $\xi$, i.e. define
$\xi$ (and therefore also $x$) as a point on a circle of length 
a multiple $m\Xi$ of $\Xi$,
$m=1,2,...$. The corresponding solution $\varphi$ of (\ref{equation})
will be defined for $x$ {\it belonging to a circle of length}
$m\Xi\sqrt{1-v^2}$, as well!

{\bf Remark 3.} The ``half-array of (anti)solitons'' solutions have no
analog in the pure sine-Gordon case.

{\bf Remark 4.}  In the list we may have to add 
for either $\alpha=0$ and $v^2<1$, or $\alpha>0$ and $v=0$,
the mentioned solutions $\varphi(x,t)=g(\pm \xi)$ such that
\be
\lim\limits_{x\to -\infty}\varphi(x,t)=-\sin^{-1}\gamma=
\lim\limits_{x\to \infty}\hat\varphi(x,t),
\ee
in case investigation should establish their stability.
They would describe rigidly bounded
soliton-antisoliton pairs moving with velocity $v$.

\subsection*{Acknowledgments}

We are indebted to C. Nappi for much information on the present
state-of-the-art of research on the Josephson effect, bibliographical
indications and stimulating discussions. It is also a pleasure to thank P.
Renno for his encouragement and comments, A. Barone and R. Fedele for
their useful suggestions.

\end{document}